# Deduction of Initial Strategy Distributions of Agents in Mix-game Model

Chengling Gou,

Physics Department, Beijing University of Aeronautics and Astronautics
37 Xueyuan Road, Heidian District, Beijing, 100083, China
gouchengling@buaa.edu.cn , gouchengling@hotmail.com

Abstract: This paper reports the effort of deducing the initial strategy distributions of agents in mix-game model which is used to predict a real financial time series generated from a target financial market. Using mix-game to predict Shanghai Index, we find the time series of prediction accurate rates is sensitive to the initial strategy distributions of agents in group 2 who play minority game, but less sensitive to the initial strategy distributions of agents in group 1 who play majority game. And agents in group 2 tend to cluster in full strategy space (FSS) if the real financial time series has obvious tendency (upward or downward), otherwise they tend to scatter in FSS. We also find that the initial strategy distributions and the number of agents in group 1 influence the level of prediction accurate rates. Finally, this paper gives suggestion about further research.

## 1. Introduction

Forecasting financial markets is a continuous effort of researchers, and attracts great attention of practitioners. The use of agent-based models to forecast financial markets is a new attempt in this area. Neil F. Johnson etc. reported on a technique based on multi-agent games which has potential use in predicting future movements of financial time-series. In their articles, minority game (MG) [1] is trained on a real financial time series, and then run into the future to extract next step and multi-step predictions [2, 3, 4]. In order to improve the forecasting accurate rate, C. Gou used mix-game [5, 6] to predict Shanghai Index. She found that using mix-game can improve forecasting accurate rate at least 3% more than using minority game and the forecasting accurate rate is sensitive to the initial strategy distributions of agents [7]. The initial strategy distribution (ISD) is distributed randomly in full strategy space and kept unchangeable during the game. Since the environment of a financial market is changeable, agents in the market need to adjust their strategies to survive in the market. For the purpose of prediction, we also need to tune the ISD of mix-game model so as to get a good accurate rate of prediction, i.e. agents in mix-game model need to change their strategies. Therefore, understanding what factors influence the accurate rates of prediction and how agents in mix-game move in their strategy spaces is important for predicting a financial time series. In order to do so, this paper looks at the backward question that we try to deduce the strategy distributions of agents in their strategy spaces in mix-game according to a real financial time series.

This paper is organized as following. Section 2 introduces the model and the methodology of deducing the strategy distributions of agents. Section 3 reports the results and discussion. Section 4 gives the conclusion.

## 2. The model and the methodology

### 2.1 Mix-game model

Since mix-game model is an extension of minority game [1], its structure is similar to MG. In mix-game, there are two groups of agents; group 1 plays the majority game, and group 2 plays the minority game. N (odd number) is the total number of the agents, and N1 is number of agents in group 1. The system resource is R=0.5*N. All agents compete in the system for the limited resource R. $T_1$ and $T_2$ are the time horizon lengths of the two groups, and $m_1$ and $m_2$ denote the history memories of the two groups, respectively.

The global information only available to agents is a common bit-string "memory" of the m1 or m2 most recent competition outcomes (1 or 0). A strategy consists of a response, i.e., 0 (sell) or 1 (buy), to each possible history bit string; hence there are $2^{2^{m1}}$ or $2^{2^{m2}}$ possible strategies for group 1 or group 2, respectively, which form full strategy spaces (FSS). At the beginning of the game, each agent is assigned *s* strategies and keeps them unchangeable during the game. After each turn, agents assign one virtual point to a strategy which would have predicted the correct outcome. For agents in group 1, they reward their strategies one point if they are in the majority side. In contrast, for agents in group 2, they reward their strategies one point if they are in the minority side. Agents collect the virtual points for their strategies over the time horizon $T_1$ or $T_2$, and they use their strategies which have the highest virtual point in each turn. If there are two strategies which have the highest virtual point, agents use coin toss to decide which strategy to be used. Excess demand is equal to the number of ones (buy) which agents choose minus the number of zeros (sell) which agents choose. According to a widely accepted assumption that excess demand exerts a force on the price of the asset and the change of price is proportion to the excess demand in a financial market, the time series of the price can be calculated based on the time series of the excess demand [8, 9].

### 2.2 Prediction method

We use mix-game model to do one-step direction prediction about a real financial time series generated by a target financial market, whose dynamics are well-described by a mix-game model for a unknown parameter configuration of $T_1$, $T_2$, $m_1$, $m_2$, $N$, $N_1$ and an unknown specific realization of initial strategy choices (ISD). After we select the parameter configuration for mix-game model, we use it to predict the time series generated by the target financial market. Fig.1 shows the work chart for prediction.

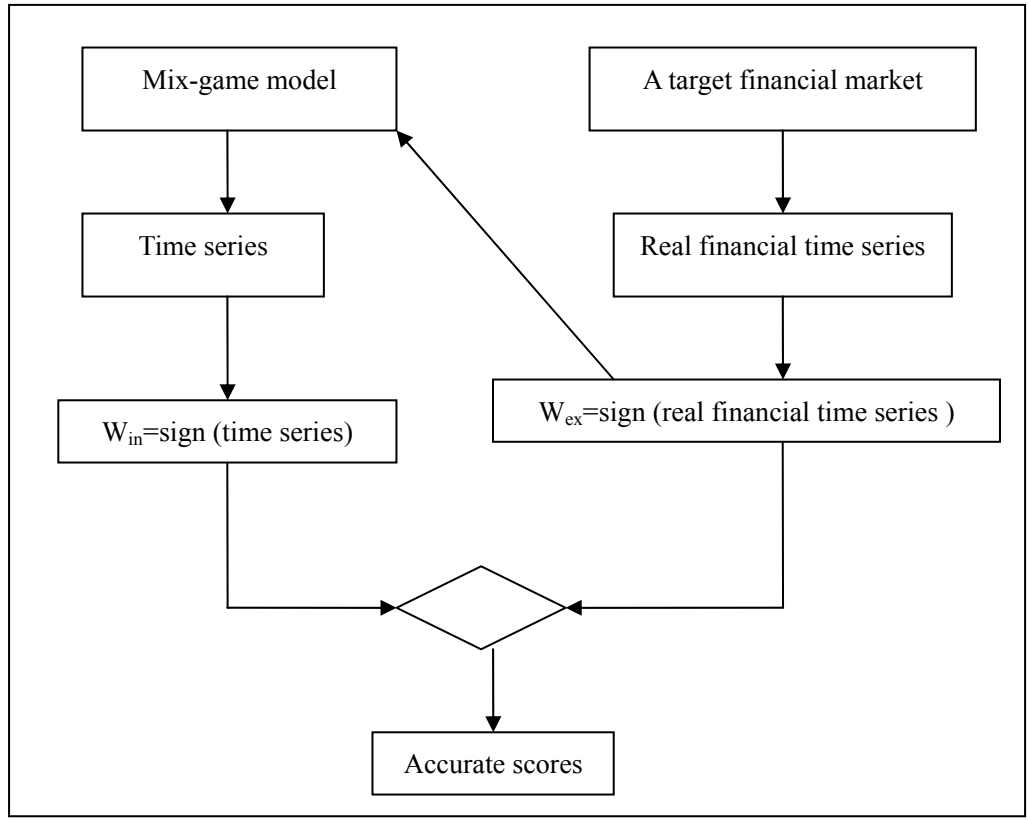

Fig.1 work chart for prediction

In the mix-game model, agents receive the global information which is strings of 1's and 0's ($W_{ex}$) digitized from the real financial time series produced by the target financial market other than that ($W_{in}$) generated by the mix-game model itself, and agents also reward their strategies according to the real financial time series ($W_{ex}$). In order to see the performances of the mix-game model, we compare the time series generated by mix-game with the real financial time series produced by the target financial market, i.e. we compare $W_{in}$ with $W_{ex}$ since we just predict the directions. If they match, the mix-game model gets one point; if not, it gets zero. This point is referred to as accurate scores. Before prediction, we use real market data of time-step $T_3$ to train the mix-game model. We calculate the accurate rate at every time-step according to the following formula:

*Accurate rate (current time) = accurate scores (current time, counted within $T_4$)/ $T_4$*     (1),

where $T_4$ is a time window within which we count the accurate scores and this time window moves along the real financial time series so that we can get a time series of prediction accurate rates. After finishing the total predicting turns, we calculate the accumulated accurate rate which is equal to total accurate scores divided by total predicting turns, i.e.

$$\textit{Accumulated accurate rate = total accurate scores/ total prediction turns} \qquad (2).$$

## 2.3 Method for deduction of strategy distributions of agents

First we need to select a proper parameter configuration for modeling the target financial market. According to reference [5], the mix-game model with some parameter configurations reproduces the stylized features of financial time series. Therefore, we need to choose parameters of $m_1$, $m_2$, $T_1$, $T_2$, N and $N_1$ when using mix-game to model financial markets. To do so, the following aspects need to be considered.

- First make sure the time series of mix-game can reproduce the stylized facts of price time series of financial markets by choosing proper parameters: $m_1<m_2=6$, $T_1<T_2$, $N_1/N<0.5$;
- Second pay attention to the fluctuation of local volatilities, and ensure that the median of local volatilities of mix-game is similar to that of the target time series;
- Third make sure the log-log plot of absolute returns look similar.

Second, we divide the time series of the real financial time series into several pieces. For each piece of time series, we use simple generic algorithm to search for the best ISD of agents by which the model can give the best prediction for this piece of real financial time series [7]. Therefore, we can see how the strategy distributions of agents change along the real financial time series. Usually we need to adjust $T_1$, $T_2$ and $N_1$ in order to get high prediction accurate rate during this process.

## 3. Results and discussions

As an example, we use the above methodology to deduce the strategy distributions of agents in mix-game which is used to game to predict Shanghai Index dating from 02-07-2002 to 19-03-2004, as shown in Fig.2. We can see that there are downward tendency, upward tendency and balance states in the time series shown in Fig.2. So this period of Shanghai Index is a typical example for study. According to reference [5], we get one configuration of parameters of mix-game for simulation of Shanghai Index which is $m_1=3$, $m_2=6$, $T_1=12$, $T_2=60$, $N=201$, $N_1=40$, $s=2$.

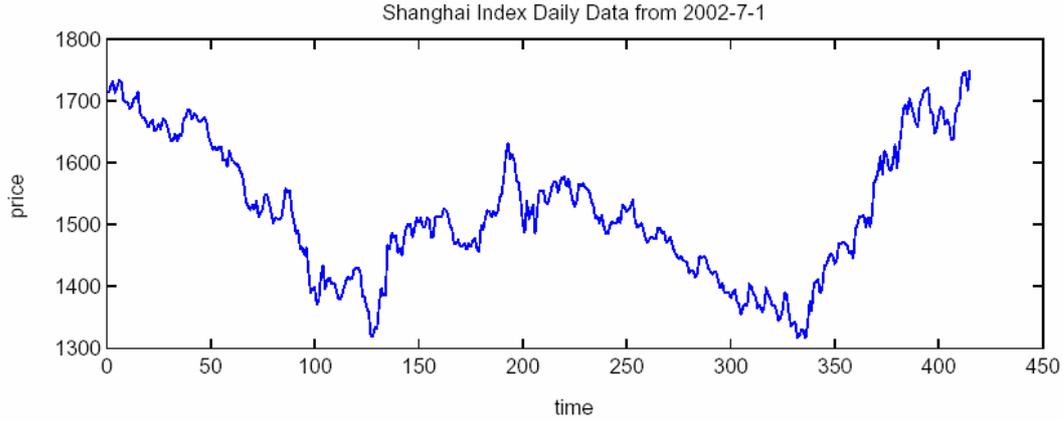

Fig.2 Shanghai Index daily data from 02-07-2002 to 19-03-2004.

A strategy is a response, i.e., 0 (sell) or 1 (buy), to each possible history bit string. Hence, we can turn the strategy into an integer value and plot it, and the integer value goes from zero to ($2^{2^{m1}}-1$) or ($2^{2^{m2}}-1$) for group 1 or group 2, respectively. Since each agent has 2 strategies, it is possible to illustrate ISD in a 2D-figure, as shown in Fig.3a where x-axis represents 'strategy R' and y-axis represents 'strategy R'' so that any point in this figure represents 2 strategies which one agent holds and all points in this figure form the initial strategy distribution（ISD）. Following the method as mentioned in section 2, we deduce the ISDs of the mix-game model according to the real financial time series shown in Fig.2. By analyzing these ISDs, we can get to know what factors influence the accurate rates of prediction and how agents move in their strategy spaces.

**3.1 What factors influence the accurate rates of prediction?**

In reference [7], C. Gou found that prediction accurate rates are sensitive to ISD. However, there are two ISDs in mix-game: $ISD_1$ is the strategy distribution of group 1; $ISD_2$ is the strategy distribution of group 2. So the question is the prediction accurate rate is sensitive to both $ISD_1$ and $ISD_2$ or just any of them. To look at this issue, we did simulations with three different $ISD_1$ but the same $ISD_2\_a$ in which all agents have the same strategies. Fig.3a, Fig.4a and Fig.5a show the three different $ISD_1$ with $N_1=90$, $m_1=3$: all agents in group 1 have different strategies in $ISD_1\_a\_1$ (Fig.3a); all agents in group 1 have the same strategies in $ISD_1\_a\_2$ (Fig.4a); at least two agents in group 1 have the same strategies in $ISD_1\_a\_3$ (Fig.5a).

The time series of prediction accurate rates of mix-game are shown in Fig.3b for initial strategy distributions of $ISD_1\_a\_1$ and $ISD_2\_a$, Fig.4b for initial strategy distributions of $ISD_1\_a\_2$ and $ISD_2\_a$ and Fig.5b for initial strategy distributions of $ISD_1\_a\_3$ and $ISD_2\_a$, respectively. Comparing these three figures, we can find these three time series of the prediction accurate rates of mix-game for Shanghai Index dating from 02-07-2002 to 19-03-2004 look similar despite of the obvious difference of their $ISD_1$. This implies that the time series of prediction accurate rates is mainly influenced by $ISD_2$. That is to say, the time series of prediction accurate rates are more sensitive to $ISD_2$, but less sensitive to $ISD_1$. A possible explanation for this observation is that the characteristics of minority game dominants in mix-game if $N_1/N<0.5$ and agents in group 1 are just "trend followers" [6].

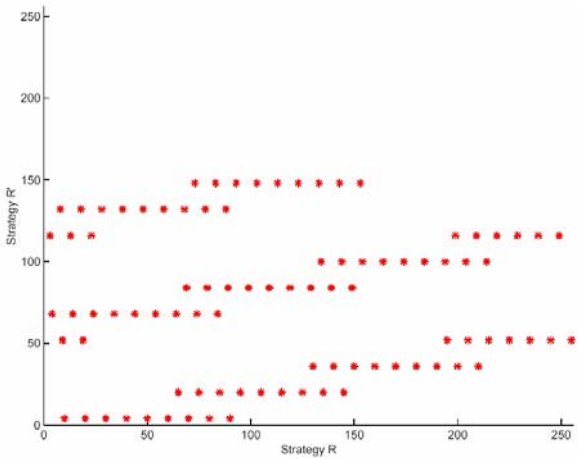 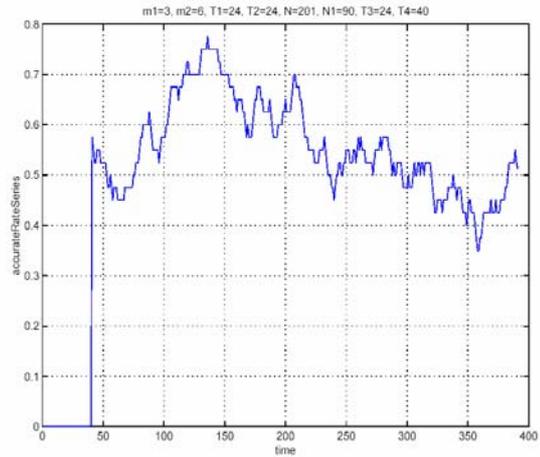

Fig.3a　　　　　　　　　　　　　　　　　　　　Fig.3b

Fig.3　Fig.3a shows $ISD_1\_a\_1$ with $N_1=90$, $m_1=3$. Fig.3b shows the time series of prediction accurate rates of mix-game with $ISD_1\_a\_1$ and $ISD_2\_a$.

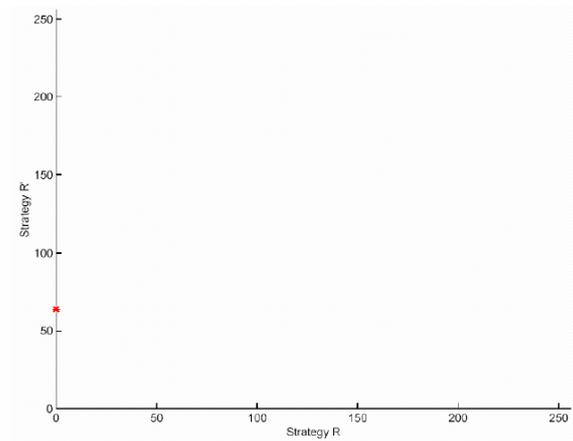 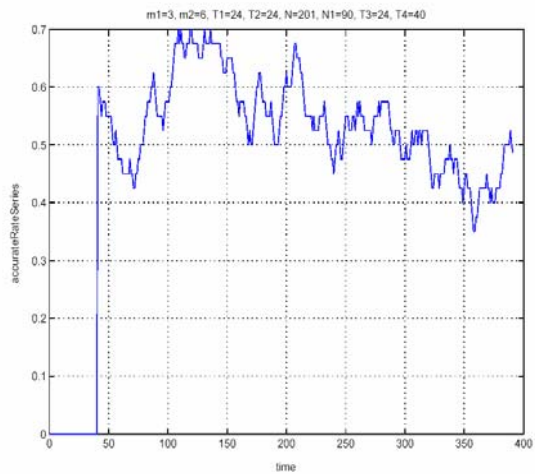

Fig.4a　　　　　　　　　　　　　　　　　　　　Fig.4b

Fig.4　Fig.4a shows $ISD_1\_a\_2$ with $N_1=90$, $m_1=3$. Fig.4b shows the time series of prediction accurate rates of mix-game with $ISD_1\_a\_2$ and $ISD_2\_a$.

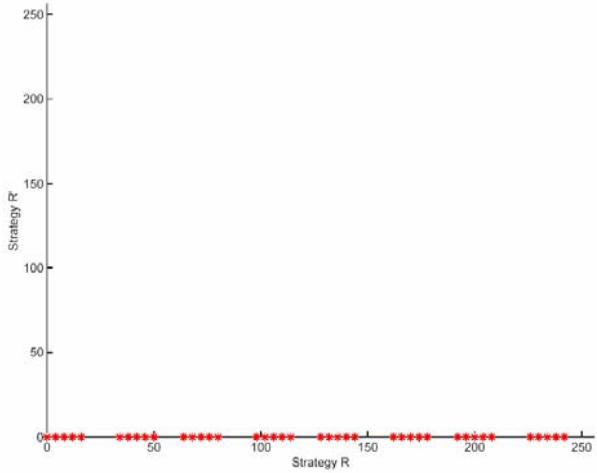 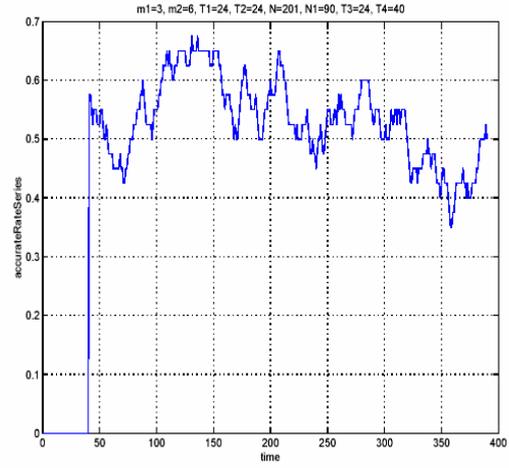

| Fig.5a | Fig.5b |

Fig.5  Fig.5a shows $ISD_1\_a\_3$ *wit* $N_1=90$, $m_1=3$. Fig.5b shows the time series of prediction accurate rates of mix-game with $ISD_1\_a\_3$ and $ISD_2\_a$.

However, we also find that $ISD_1$ have influence on the accurate rate level, as shown in table 1 which shows the accumulated accurate rates of mix-game at 390 predicting turns with $ISD_2\_a$ and three different $ISD_1$ ($ISD_1\_a\_1$, $ISD_1\_a\_2$ and $ISD_1\_a\_3$). From table 1, we notice that the accumulated accurate rates are different with different $ISD_1$. This implies that $ISD_1$ can influence the level of the accumulated accurate rates.

Table 1  Accumulated accurate rates of mix-game at 390 predicting turns with $ISD_2\_a$ and three different $ISD_1$.

| $ISD_1$ | $ISD_1\_a\_1$ | $ISD_1\_a\_2$ | $ISD_1\_a\_3$ |
|---|---|---|---|
| Accumulated accurate rate | 56.0% | 54.5% | 53.7% |

Table 2 shows the accumulated accurate rates with different $N_1$, where $m_1=3$, $m_2=6$, $N=201$, $T_1=T_2=T_3=24$, $s=2$, $ISD_1$ similar to $ISD_1\_a\_1$ and $ISD_2$ similar to $ISD_2\_a$ at 390 predicting turns. From table 2, we can see that the number of agents in group 1 greatly influences the accumulated accurate rates and the accumulated accurate rates increase with the increase of the number of agents in group 1. A possible explanation for this finding is that mix-game can simulate Shanghai Index better than MG [5].

Table 2  Accumulated accurate rates with different $N_1$, where $m_1=3$, $m_2=6$, $N=201$, $T_1=T_2=T_3=24$, $s=2$, $ISD_1$ similar to $ISD_1\_a\_1$ and $ISD_2$ similar to $ISD_2\_a$ at 390 predicting turns

| N1 | 0 | 10 | 40 | 90 |
|---|---|---|---|---|
| Accumulated accurate rate | 48.0% | 53.7% | 55.0% | 56.0% |

**3.2 Agents move in their strategy spaces**

Now we look at how agents in mix-game move in their strategy spaces while we use the mix-game to predict Shanghai Index as shown in Fig.2. First we divide this time series into four pieces: the first piece from 0 to 126th time-step which has the downward tendency; the second piece from 127th to 230th time-step which is characteristic of two balance states; the third piece from 231st to 341st time-step which has the downward tendency; the 4th piece from 342nd to 415th time-steps which has the upward tendency. Then we use the method mentioned in section 2 to get the best ISDs for the above pieces of Shanghai Index so that we can see how agents move in their strategy spaces. Since the time series of prediction accurate rates is only sensitive to $ISD_2$, we focus on agents in group 2 in the following discussion.

From Fig.3b, we can see that $ISD_2\_a$ has high prediction accurate rates from 60th time-step to 150th time-step. From Fig.6b, we observe that $ISD_2\_b$ has high prediction accurate rates from 150th time-step to 200th time-step, and from 250th time-step to 300th time-step. Fig.7b shows that $ISD_2\_c$ has high prediction accurate rates from 200th time-step to 250th time-step. Fig.8b shows that $ISD_2\_d$ has high prediction accurate rates from 300th time-step to 350th time-step. Therefore, we find that $ISD_2$ changes from $ISD_2\_a$ to $ISD_2\_b$, to $ISD_2\_c$, to $ISD_2\_b$, to $ISD_2\_d$ along the time series shown in Fig.2. By analyzing $ISD_2\_a$, $ISD_2\_b$, $ISD_2\_c$ and $ISD_2\_d$, we can notice that the distribution of agents in $ISD_2$ changes from one point to half FSS and scatters further to whole FSS, then shrinks to half FSS and shrinks further. Associating the change of $ISD_2$ with the time series shown Fig.2, we can find that agents tend to cluster in FSS if the real financial time series has obvious tendency (upward or downward), otherwise they tend to scatter in FSS and the strategies of agents are more diversified in balance states. This is accordance with the common knowledge about financial markets.

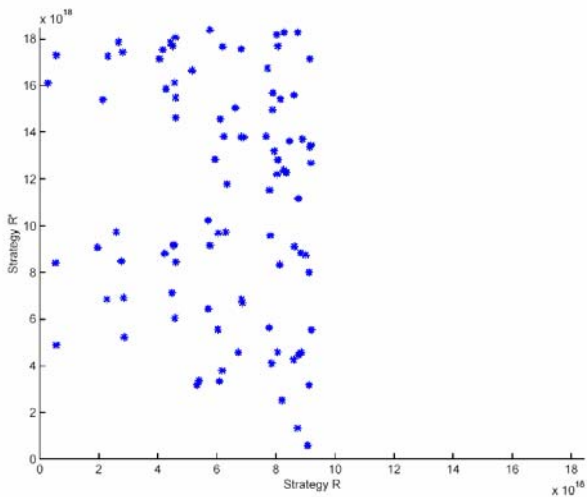
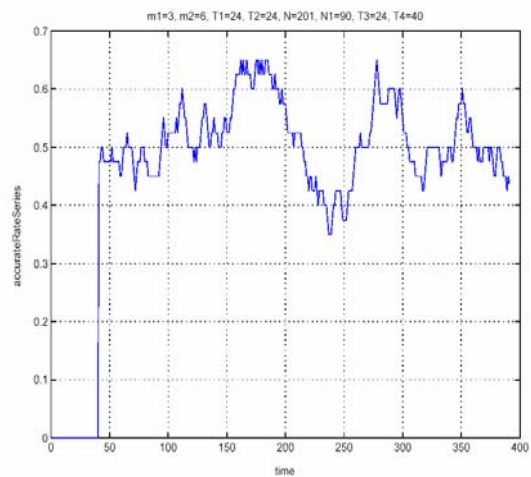

Fig.6a            Fig.6b

Fig.6    Fig.6a shows $ISD_2\_b$ and Fig.6b shows the corresponding time series of prediction accurate rates.

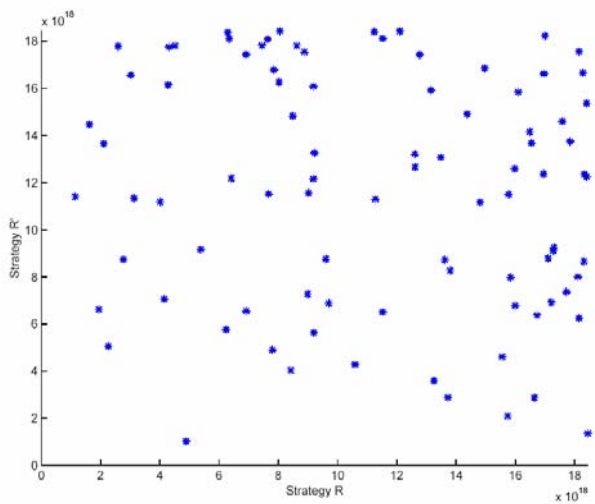 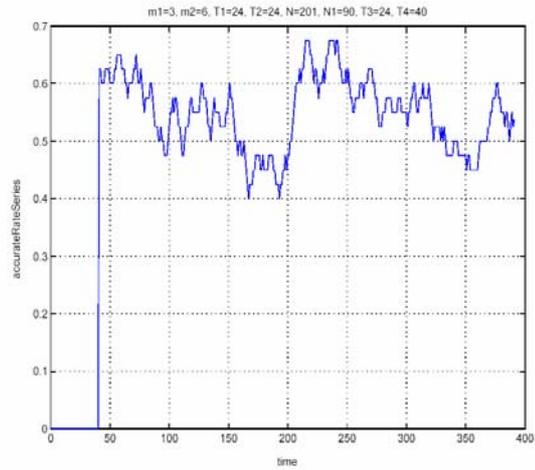

Fig.7a  Fig.7b

Fig.7  Fig.7a shows $ISD_2\_c$ and Fig.7b shows the corresponding time series of prediction accurate rates.

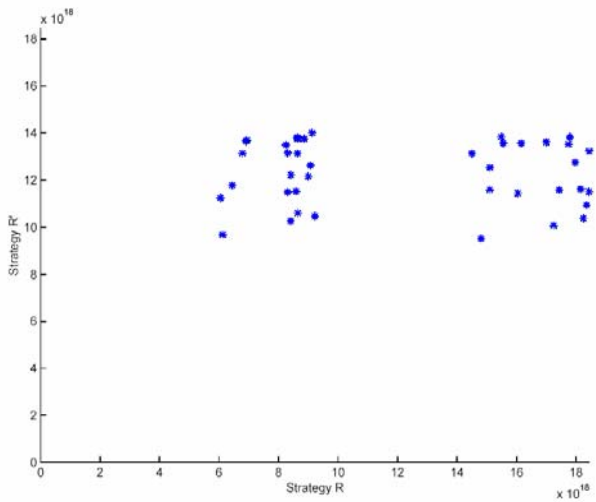 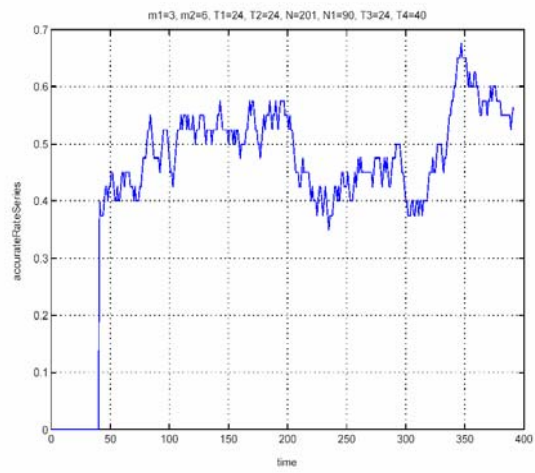

Fig.8a  Fig.8b

Fig.8 Fig.8a shows $ISD_2\_d$ and Fig.8b shows the corresponding time series of prediction accurate rates.

## 4. Conclusion

We successfully introduce a method to deduce the initial strategy distributions (ISD) of multi-agent-based mix-game model to predict a real financial time series, and can understand what factors influence the accurate rates of prediction and how agents in mix-game model move in their strategy spaces by analyzing the ISDs.

Using mix-game to predict Shanghai Index dating from 02-07-2002 to 30-12-2003, we find the time series of prediction accurate rates is more sensitive to $ISD_2$ of group 2 than $ISD_1$ of group 1, and agents in group 2 tend to cluster in FSS if the real financial time series has obvious tendency, otherwise they tend to scatter in FSS. We also find that $ISD_1$ and the number of agents in group 1 influence the level of prediction accurate rates.

The above results shed light on further research of agent-based prediction. We may let agents change their strategies during the process of prediction in order to get better prediction accurate rates by introducing generic algorithm into mix-game model. We may only let agents in group 2 to change their strategies because the prediction accurate rates are more sensitive to $ISD_2$ of group 2 than $ISD_1$ of group 1.

## Acknowledge

This research is supported by China Scholarship Council. Thanks Professor Neil F. Johnson for helpful discussion. I appreciate the referees' comments.